\documentclass[paper,notoc]{JHEP3}

\JHEPspecialurl{http://jhep.sissa.it/JOURNAL/JHEP3.tar.gz}
\usepackage{graphics}

\def\ti {\tilde}
\def\sq              {{\ti q}}
\def\st              {{\ti t}}
\def\sb              {{\ti b}}
\def\sg              {{\ti g}}
\def\rzw             {\sqrt{2}}
\def\PL              {P_L^{}}
\def\PR              {P_R^{}}
\def\Rst             {R^{\,\st}}

\def\Rsbs            {R^{\,\sb *}}
\newcommand{\nn}{\nonumber}

\newcommand{\be}{\begin{eqnarray}}
\newcommand{\ee}{\end{eqnarray}}

\newcommand{\drbar}{{\overline{\rm DR}}}
\newcommand{\msbar}{{\overline{\rm MS}}}

\newcommand{\vs}{\\ \vspace*{-5mm}}

\title{CP violation in charged Higgs decays in the MSSM}

\author{Ekaterina Christova$^{\dagger}$, Helmut Eberl$^{\ddagger}$, 
        Elena Ginina$^{\dagger}$, Walter Majerotto$^{\ddagger}$\\

    $^{\dagger}$Institute for Nuclear Research and Nuclear Energy, Sofia 1784, Bulgaria\\
    $^{\ddagger}$Institut f\"ur Hochenergiephysik der \"Osterreichischen Akademie der
     Wissenschaften,  A-1050 Vienna, Austria\\
     E-mails: \email{echristo@inrne.bas.bg}, \email{helmut@hephy.oeaw.ac.at},\\
     \hphantom{E-mails: }\email{eginina@inrne.bas.bg}, \email{majer@hephy.oeaw.ac.at}}

\abstract{In the MSSM with complex parameters loop corrections to the
decays $H^{+}\rightarrow t\,\bar{b}$ and $H^{-}\rightarrow \bar{t}\,b$
with $t \to b\, W$ and $W \to l\, \nu$ lead to CP-violating asymmetries:
a decay rate asymmetry, a forward-backward asymmetry and an energy
asymmetry. We derive explicit formulas for them and perform a detailed
numerical analysis. We study the dependence on the parameters and the
phases involved. In particular, the influence of the running Yukawa
coupling is taken into account. The decay rate asymmetry can go up
to 25\%, the forward-backward and the energy asymmetry up to 10\%.
}


\begin{document}
------------------------------------------------------------------------------------------------------------
 \tableofcontents
------------------------------------------------------------------------------------------------------------
%
%
\section{Introduction}
%
%
It is well known that supersymmetric models contain new sources of
CP violation if the parameters are complex. In the Minimal
Supersymmetric Standard Model (MSSM) the U(1) and SU(2) gaugino
mass parameters $M_1$ and $M_2$, the higgsino mass
parameter $\mu$, as well as the trilinear couplings $A_f$
(corresponding to a fermion $f$) may be complex. (Usually, $M_2$ is
made real by redefining the fields.) Non-vanishing phases of these
parameters cause CP-violating effects. While the phase of $\mu$
may be small for a supersymmetric particle spectrum of ${\cal
O}$(100 GeV) due to the experimental upper bounds of the electric
dipole moments (EDMs) of electron and neutron, the trilinear
couplings of the third generation $A_{t,b,\tau}$ are not so much
constrained and can lead to significant CP-violation \cite{nath,
phasemu}.

In the following, we study CP violation in the decays of the
charged Higgs bosons $H^{\pm}$ within the MSSM. There are three
possible decays of $H^{\pm}$ into ordinary particles: $H^{+}$\
into  $t \bar{b}$ , $\tau\nu$ and $Wh^0$ and the CP conjugated
ones, where $h^0$ is the lightest neutral Higgs boson of the MSSM.
At tree level the partial decay widths of $H^+$ and $H^-$ are
equal because of CP invariance of the Higgs potential. However,
including loop corrections with intermediate SUSY-particles, they
become different due to the CP violation induced by the complex
phases of the MSSM parameters, essentially of $A_{t,b,\tau}$.
Quite generally, these phases affect the whole Higgs sector of the
MSSM substantially \cite{pilaftsis,carenaCP}.

A full one-loop calculation within the MSSM was done of the decays
mentioned \cite{we1,we2,our,jennifer},
and the CP-violating decay rate asymmetry $\delta^{CP}=
[\Gamma_{H^+}-\Gamma_{H^-}]/[\Gamma_{H^+}+\Gamma_{H^-}]$ for these
decays was calculated. In the case of $H^{+}\rightarrow t \bar{b}$ and
$H^{-}\rightarrow \bar{t} b$ this asymmetry can go up
to $\sim$~25\%.

\begin{figure}[h!]
 \begin{center}
 \mbox{\resizebox{!}{4.3cm}{\includegraphics{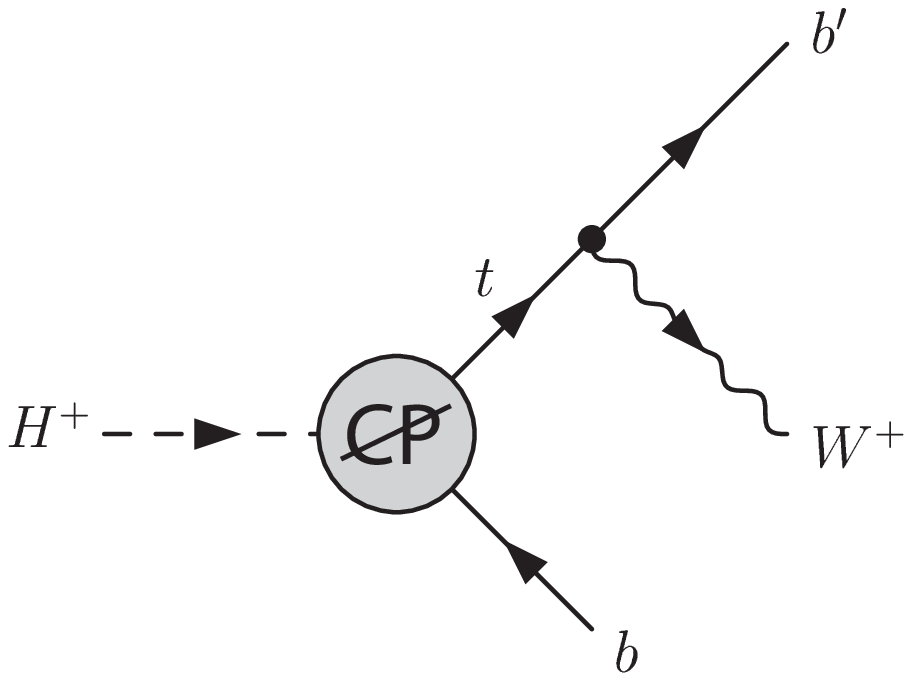}}} \hfil
 \mbox{\resizebox{!}{4.3cm}{\includegraphics{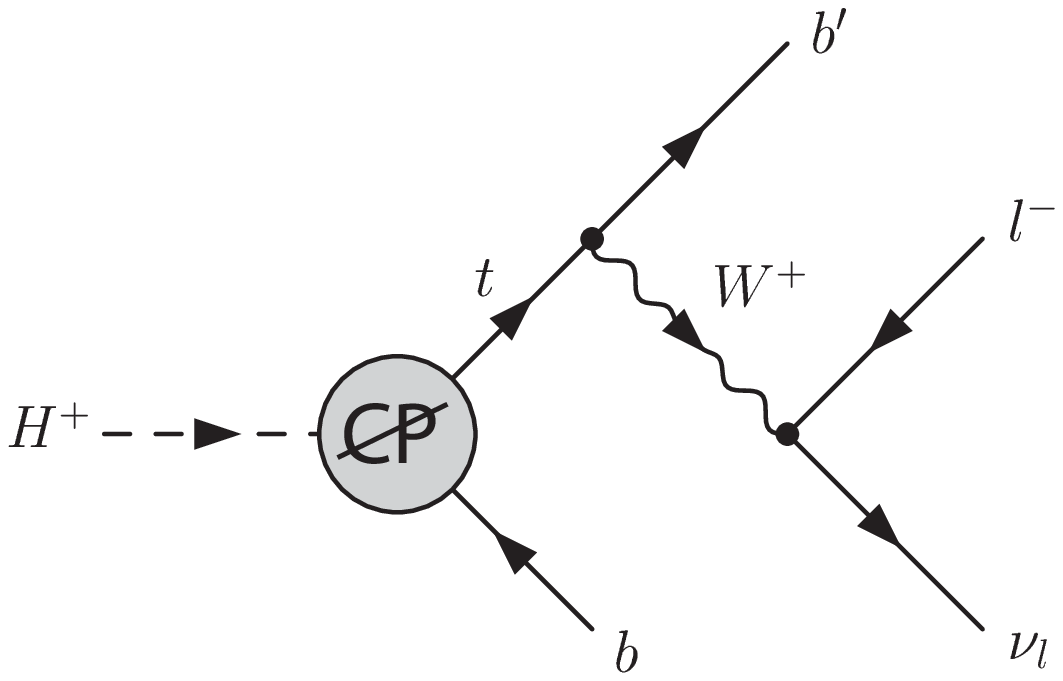}}} \vs
  \end{center}
  \caption[feynman]{The Feynman graphs of the processes we study.
  \label{fig:feyn}}
\end{figure}

In this paper, we go a step further by including the decay product
particles of the top quark, see Fig.~\ref{fig:feyn},
\[
 H^{+}\rightarrow \bar{b}\,t \rightarrow \bar{b}\,b'\,W^{+}\, ,
\]
\begin{equation}
\label{pro1} H^{-}\rightarrow b\,\bar{t} \rightarrow b\,\bar{b}'\,W^{-}\, ,
\end{equation}
and
\[
 H^{+}\rightarrow \bar{b}\,t \rightarrow \bar{b}\,b'\,W^{+} \rightarrow
\bar{b}\,b'\,l^{+}\,\nu_l \, ,
\]
\begin{equation}
\label{pro2} H^{-}\rightarrow b\,\bar{t}\rightarrow b\,\bar{b}'\,W^{-} \rightarrow
b\,\bar{b}'\,l^{-}\,\bar\nu_l\, .
\end{equation}
We only consider CP violation induced by the loop diagrams of
$H^\pm t b$ vertex. We neglect CP violation in the $t W b'$ vertex.
In the Standard model it is very small and in the MSSM 
for $m_{H^+} > m_t$ all two-body
decays of top into SUSY partners are excluded kinematically.

In particular, we will exploit the polarization of the top quark.
The top-quark decays before forming a bound state due to its large
mass, so that the polarization can be measured by the angular
distributions of its decay products. The polarization is very
sensitive to CP violation. We will consider suitable CP violating
forward-backward and energy asymmetries by using angular or energy
distributions of the decay particles. The asymmetries
also depend on the sensitivity of $b'$ in (\ref{pro1}) or the
final lepton $l^{\pm}$ in (\ref{pro2}) to the top-quark
polarization. Further we make a numerical analysis
for different values of the MSSM parameters.

This paper is organized in the following order. In Section~\ref{sec:2} we
present the formalism we use. There are subsections devoted to the
polarization of the top quark and the CP-violating asymmetries.
Sections~\ref{sec:3} and~\ref{sec:4} contain the angular and energy distributions and
the analytic results for the asymmetries of (\ref{pro1}) and
(\ref{pro2}), respectively. The numerical results are discussed in
Section~\ref{sec:5}. Section~\ref{sec:6} contains the conclusions. 
In Appendix~\ref{sec:A} the formulas used for running Yukawa couplings 
$h_b$ and $h_t$ are given. In Appendix~\ref{sec:B} we
point out an error made in an equation of \cite{we1}.
%
%
\section{Formalism} \label{sec:2}
%
%
In order to obtain the analytic expressions for the differential
partial decay rates of (\ref{pro1}) and (\ref{pro2}), we follow
the formalism of \cite{form}. In accordance with it, for both of the
processes we write
\begin{equation}
d\Gamma^{\pm}_f = d\Gamma_{H^{\pm}}d\Gamma^f_{ {t},
{\bar{t}}}{E_{t,\bar{t}}\over m_t\Gamma_{t}}\, , \quad f = b',\,l\, .
\end{equation}
$E_{t,\bar{t}}$ is the energy of $t(\bar{t})$-quark, and
$\Gamma_t$ is the total decay width of the t-quark.
$d\Gamma_{H^{\pm}}$ is the differential partial decay rate of the
process $H^{\pm}\rightarrow tb$ when CP-violation is included:
\begin{equation}
 d\Gamma_{H^{\pm}}={1 \over 2m_H
}|{\cal M}_{H^{\pm}}|^2d\Phi_{H^{\pm}}\,.
\end{equation}
where $d\Phi_{H^{\pm}}$ is the relevant phase space
element and
\begin{equation}
\label{amp1} {\cal M}_{H^{+}}=i
\bar{u}(p_t)[Y_b^{+}P_R+Y_t^+P_L]u(-p_{\bar b})\,,
\end{equation}
\begin{equation}
\label{amp2} {\cal M}_{H^{-}}=i
\bar{u}(p_b)[Y_t^{-}P_R+Y_b^-P_L]u(-p_{\bar t})\,,
\end{equation}
\begin{equation}
\label{del1} Y_t^{\pm}=y_t+\delta Y^{ \pm }_t, \qquad
Y_b^{\pm}=y_b+\delta Y^{ \pm }_b\,.
\end{equation}
Here $y_t$ and $y_b$ are
the $\overline{\rm DR}$ running couplings, see Appendix~\ref{sec:A},
$\delta Y^{\pm}_{t,b}$ are the SUSY-induced loop corrections, which
most generally have CP-invariant and CP-violating parts:
\begin{equation}
\label{del2}
 \delta Y^{\pm}_{t,b}=\delta Y^{inv}_{t,b}\pm {1\over 2}\delta
 Y^{CP}_{t,b}\, .
\end{equation}

 $d\Gamma^f_{ {t},{\bar{t}}}$ is the  differential partial rate of the process $t
\rightarrow  b' W^{\pm}$ or $t\rightarrow  b' l^{\pm} \nu$ when the t-quark is
polarized and its polarization is determined in the former process $H^{\pm}\rightarrow t\bar{b}$:
\begin{equation}
d\Gamma_{ {t}, {\bar{t}}}^f=\Gamma_f^0\bigg[1\pm \alpha_f
m_t{(\xi_{t,\bar{t}}p_f)\over (p_{t }p_f)}\bigg] d\Phi_{ {t},
{\bar{t}}}^f\, .
\end{equation}
${\xi}_{t}^{\alpha}$ is the polarization vector of the t-quark and
$ d\Phi_{ {t}, {\bar{t}}}^f$ are the phase space elements. The index
$f$ stands for the corresponding fermion
($f=b',l$) and $\alpha_f$ determines its sensitivity to the
polarization of the t-quark:
\begin{equation}
\alpha_b={m_t^2-2m_W^2\over m_t^2+2m_W^2}, \qquad \alpha_l=1\,.
\end{equation}
In the kinematics of both of the processes (\ref{pro1}) and
(\ref{pro2}) we work in the approximation
\begin{equation}
m_l^2/m_W^2\simeq m_b^2/m_t^2\simeq m_b^2/m_W^2\simeq 0\,,
\end{equation}
but we keep $m_b\neq 0$ in the couplings, where it is multiplied
by $\tan\beta$.
%
%
\subsection{The t-quark polarization vector}
%
%
The polarization four-vectors $ {\xi}_{t}^{\alpha} $ and $
{\xi}_{\bar{t}}^{\alpha} $ for the considered processes are
covariantly given by the expressions \cite{form}:
\begin{equation}
\xi_t ^{\alpha}=\bigg(g^{\alpha \beta
}-{p_t^{\alpha}p_t^{\beta}\over m_t^2}\bigg){Tr[M_{H^+}
(-\Lambda(- p_{\bar{b}}))\overline{M}_{H^+}
 \Lambda( p_t)\gamma_{\beta}\gamma_5]\over Tr[M_{H^+} (-1)\Lambda(-
p_{\bar{b}})\overline{M}_{H^+}
 \Lambda( p_t)]}\,,
\end{equation}
\begin{equation}
\xi_{\bar{t}}^{\alpha}=\bigg(g^{\alpha \beta
}-{p_{\bar{t}}^{\alpha}p_{ \bar{t}}^{\beta}\over
m_t^2}\bigg){Tr[M_{H^-}(-\Lambda(- p_{\bar{t}}))\overline{M}_{H^-}
 \Lambda( p_{b})\gamma_{\beta}\gamma_5]\over Tr[M_{H^-}(-1)\Lambda(-
p_{\bar{t}})\overline{M}_{H^-}
 \Lambda( p_{b})]}\,,
\end{equation}
where
\begin{equation}
M_{H^+}=Y_b^{+}P_R+Y_t^+P_L, \qquad M_{H^-}=Y_t^{-}P_R+Y_b^-P_L\,,
\end{equation}
\begin{equation}
\overline{M}=\gamma_0M^\dagger\gamma_0, \qquad  \Lambda (
p_{t})=\slash\!\!\!{p}_t+m_t\,.
\end{equation}
Thus we obtain:
\begin{equation}
\label{polvec} \xi_{t,\bar{t}}^{\alpha}=m_t{\cal
P}^{\pm}Q_{b,\bar{b}}^{\alpha}, \qquad Q_{b,\bar{b}}^{\alpha}=
p_{\bar{b},b}^{\alpha}-{(p_{t }p_{\bar{b} })\over
m_t^2}p_{t,\bar{t}}^{\alpha}\,.
\end{equation}
Notice that the four-vector $Q_{b,\bar{b}}^{\alpha}$ is the only
four-vector in $H^\pm \to tb$ that can be
constructed so that it satisfies the orthogonal condition $(\xi_t
p_t)=0$. The polarization vectors (\ref{polvec}) have CP-invariant
and CP-violating parts, and the CP-violating parts are only due to
the loop corrections:
\begin{equation}
{\cal P}^{\pm}=  \pm{\cal P}^{inv}+{\cal P}^{CP}\,,
\label{eq:Ppm}
\end{equation}
\begin{equation}
{\cal P}^{inv}= {y_t^2 - y_b^2 \over  (y_t^2 + y_b^2 ) (p_{t
}p_{\bar{b} })- 2 m_t m_b y_t y_b }\,,
\end{equation}
\begin{equation}
{\cal P}^{CP}=  {y_t {\rm Re}(\delta Y_t^{CP})-  y_b {\rm Re}(\delta Y_b^{CP})\over
(y_t^2 + y_b^2) (p_{t }p_{\bar{b}})- 2 m_t m_b y_t y_b} \, .
\label{PCP}
\end{equation}
The explicit forms of the individual contributions to  ${\rm Re} (
\delta Y^{CP}_{t,b}$) are taken from \cite{we1}.

\subsection{CP-violating asymmetries}
%
%
The CP-violating decay rate asymmetry $\delta ^{CP}$ is given by
the expression
\begin{equation}
  \delta ^{CP}={\Gamma_+ -\Gamma_- \over \Gamma_+ +\Gamma_-} \, .
\label{eq:dCPdef}
\end{equation}
In (\ref{eq:dCPdef}) $\Gamma_\pm$ are the partial decay widths of $H^\pm$.\\
Next, we construct a CP-violating forward-backward (FB) asymmetry
$\Delta A^{CP}$ from the FB asymmetries $ A^{FB}_{\pm}$ using the
angular distributions of the processes
\begin{equation}
\Delta A^{CP}= A^{FB}_+ -  A^{FB}_- \, ,
\end{equation}
where
\begin{equation}
 A^{FB}_{\pm}= {\Gamma_{\pm}^F-\Gamma_{\pm}^B \over
 \Gamma_{\pm}^F+\Gamma_{\pm}^B}\, ,
\end{equation}
\begin{equation}
\Gamma_{\pm}^F=\int_0^{\pi\over 2} {d\Gamma^{\pm}\over d\cos
\theta}d\cos \theta  \quad {\rm and}\quad
\Gamma_{\pm}^B=\int_{\pi\over 2}^{\pi}{d\Gamma^{\pm}\over d\cos
\theta}d\cos \theta \, ,
\end{equation}
i.e. $\Gamma_{\pm}^F$ are the number of particles
/antiparticles measured in the forward direction of the decaying
$t$/$\bar t$ quarks, etc.

Analogously, a CP-violating energy
asymmetry $\Delta R^{CP}$ can be defined, using the energy
distributions of the processes
\begin{equation}
\Delta R^{CP}=R_+ -  R_- \, ,
\end{equation}
where $R_{\pm}$ are
\begin{equation}
\label{eq:R}
 R_{\pm}={\Gamma_{\pm} (x>x_0)-\Gamma_{\pm}(x<x_0)
\over \Gamma_{\pm}(x>x_0)+\Gamma_{\pm}(x<x_0)} \, .
\end{equation}
$x$ is a dimensionless variable proportional to the energy,
and $x_0$ is any fixed value in the energy interval.
%
%
\section{The \boldmath $H^{\pm}\rightarrow W^{\pm}bb'$ process} \label{sec:3}
%
%
Following the formalism of Section 2 for the differential partial decay rate of
the process (\ref{pro1}) in the rest frame of $H^{\pm}$, we obtain
\begin{equation}
d\Gamma^{\pm}_b= \Gamma_{H^{\pm}} \Gamma_{b}^0\bigg[1\pm \alpha_b
m_t{(\xi_{t,\bar{t}}p_{b',\bar{b}'})\over (p_{t }p_{b'
})}\bigg]{E_{t,\bar{t}}\over m_t\Gamma_{t}}
d\Phi_{b',\bar{b}'} \, .
\end{equation}
$\Gamma_{H^{\pm}}$ is the partial decay width of the process
$H^{\pm}\rightarrow tb$, assuming CP-violation in its vertex
\begin{equation}
\Gamma_{H^\pm} = {\cal C}_H(\Gamma^{inv}\pm \Gamma^{CP})\,,
\label{eq:GammaHpm}
\end{equation}
where
\begin{equation}
{\cal C}_H={3\alpha_{\omega}\lambda^{1/2}(m_H^2,m_t^2,m_b^2) \over
4m_H^3m_W^2 }, \qquad \alpha_{\omega}={g^2\over 4\pi}\,,
\end{equation}
\begin{equation}
\Gamma^{inv}= (y_t^2 + y_b^2)(p_tp_{\bar{b}})-2m_t m_b y_t y_b \,,
\end{equation}
\begin{equation}
\Gamma^{CP}=   \left[ y_t {\rm Re}(\delta Y_t^{CP})+  (y_b
{\rm Re}(\delta Y_b^{CP}) \right](p_tp_{\bar{b}})-
m_t m_b \left[ y_t{\rm Re} (\delta Y_b^{CP})+ y_b {\rm Re}(\delta Y_t^{CP}) \right]\, ,
\label{GammaCP}
\end{equation}
\begin{equation}
\lambda (x,y,z)=x^2+y^2+z^2-2xy-2xz-2yz, \quad
\lambda^{1/2}(m_H^2,m_t^2,m_b^2)\approx m_H^2-m_t^2\, ,
\end{equation}
\begin{equation}
(p_tp_{\bar{b}})= {1\over 2}(m_H^2-m_t^2-m_b^2)\simeq {1\over 2}(m_H^2-m_t^2 )\, ,
\end{equation}
\begin{equation}
\Gamma_{b }^0={g^2(m_t^2-m_W^2)(m_t^2+2m_W^2)\over 8 m_W^2 E_{t}
}\, , \quad  {\rm and}\quad  d\Phi_{b',\bar{b}'}= -{(m_t^2-m_W^2) d\cos\theta_{b',\bar{b}'} \over
16 \pi E_{t }^2 (1-\beta_{t }\cos\theta_{b',\bar{b}'})^2}\, .
\end{equation}
%
%
\subsection{Angular distributions}
%
%
For the angular distributions of $b'(\bar{b}')$ when the
3-momentum of the $t(\bar t)$-quark is along the z-axis one gets
\begin{eqnarray}
{d\Gamma^{\pm}_b\over d\cos\theta_{b',\bar{b}'}} = {{\cal C}_{b}
\over (1-\beta_{t }\cos \theta_{b',\bar{b}'})^2 }\bigg\{
\Gamma^{inv} \pm \Gamma^{CP}+  \hspace{6cm} \\ \nonumber
   \alpha_b m_t^2 [\Gamma^{inv}{\cal P}^{inv} \pm (\Gamma^{CP}{\cal
P}^{inv} +\Gamma^{inv}{\cal P}^{CP})] \bigg({E_{\bar b}(1+\cos
\theta_{b',\bar{b}'})\over E_{t }(1-\beta_{t }\cos
\theta_{b',\bar{b}'})}-{(p_{t }p_{\bar{b}})\over m_t^2}
\bigg)\bigg\}\,,
\end{eqnarray}
where
\begin{equation}
{\cal C}_{b}=-{3\alpha_{\omega}^2 |\vec{p}_{t} |
  (m_t^2-m_W^2)^2(m_t^2+2m_W^2)\over 64 m_H^2 m_W^4 E_{t }^2m_t\Gamma_t }\,,
\end{equation}
\begin{equation}
  \beta_{t }={|\vec{p}_{t
}|\over E_{t }}, \quad |\vec{p}_{t} | ={\lambda^{1/2}(m_H^2,
m_t^2, m_b^2)\over 2m_H}\simeq {m_H^2-m_t^2\over 2m_H}\,,
\end{equation}
\begin{equation}
 E_{t}= {m_H^2+m_t^2-m_b^2\over 2m_H} \simeq {m_H^2+m_t^2 \over
 2m_H}, \quad E_{\bar b}={m_H^2+m_b^2-m_t^2\over 2 m_H}\simeq {m_H^2
-m_t^2\over 2 m_H}\, .
\end{equation}
We are interested in the CP-violating contributions to the loop
corrections of the $H^{\pm}bt$ vertex (\ref{del2}). The quantities
$ \delta Y_t^{CP} $ and $ \delta Y_b^{CP}$ enter the two
independent combinations $\Gamma^{CP}$ and ${\cal P}^{CP}$.One
therefore needs two measurements to determine them. The decay rate
asymmetry $\delta_b ^{CP}$ for process (\ref{pro1}) measures $\Gamma^{CP}$ given
in (\ref{GammaCP}),
\begin{equation}
  \delta_b ^{CP} =\frac{N_{b'} - N_{\bar b'}}{N_{b'} + N_{\bar b'}} ={\Gamma^{CP}\over
\Gamma^{inv}}\,,
\end{equation}
where $N_{b'(\bar b')}$ are the total number of  $b'$ ($\bar b'$) in $H^\pm \to bb'W^\pm$ decay.

The CP-violating angular asymmetry $\Delta A^{CP}$ measures the other
combination ${\cal P}^{CP}$ given in (\ref{PCP}). We have
\begin{equation}
\label{eq:DeltaAbCP}
\Delta A^{CP}_b=2\alpha_bm_t^2m_H^2{m_H^2-m_t^2 \over(m_H^2+m_t^2)^2}{\cal P}^{CP}
\end{equation}
The FB asymmetries are given by
\begin{equation}
A^{FB}_{b\,\pm}=\beta_t+\alpha_b
m_t^2m_H^2{(m_H^2-m_t^2)\over(m_H^2+m_t^2)^2}{(\Gamma^{inv}{\cal
P}^{inv}\pm \Gamma^{CP}{\cal P}^{inv}\pm
  \Gamma^{inv}{\cal P}^{CP})\over
 \Gamma^{inv}\pm \Gamma^{CP}} \, .
\label{eq:AbFBall}
\end{equation}
Using the expansion
\begin{equation}
{(\Gamma^{inv}{\cal
P}^{inv}\pm \Gamma^{CP}{\cal P}^{inv}\pm
  \Gamma^{inv}{\cal P}^{CP})\over
 \Gamma^{inv}\pm \Gamma^{CP}} = {\cal P}^{inv}\pm {\cal P}^{CP} + {\rm higher\,orders}\, ,
\label{eq:P1loopexpand}
\end{equation}
we get at one-loop level
\begin{equation}
A^{FB}_{b\,\pm}=\beta_t+\alpha_b
m_t^2m_H^2{(m_H^2-m_t^2)\over(m_H^2+m_t^2)^2}
\left({\cal P}^{inv}\pm {\cal P}^{CP}\right) \, .
\label{eq:AbFB}
\end{equation}
%
%
\subsection{Energy distributions}
%
%
We write the energy distribution of $b' (\bar{b}')$ as a function
of $ x={E_{b'}/ m_H}$ ($ {x}={E_{\bar{b}'}/ m_H} $):
\begin{equation}
 {d\Gamma^{\pm}\over dx } ={\cal
 C}_{E } [c_0^{\pm}+c_1^{\pm}x ],
\end{equation}
where
\begin{equation}
{\cal C}_{E }={3\alpha_{\omega}^2 (m_t^2-m_W^2)(m_t^2+2m_W^2)\over
2^6 m_W^4m_H^2 m_t\Gamma_t},\quad c_0^{\pm}=c_0^{inv}\pm
c_0^{CP},\quad c_1^{\pm}=c_1^{inv}\pm c_1^{CP},
\end{equation}
\begin{equation}
c_0^{inv}=\Gamma^{inv}-\alpha_b{(m_H^2+m_t^2)\over
2}\Gamma^{inv}{\cal P}^{inv},
\end{equation}
\begin{equation}
c_0^{CP}=\Gamma^{CP}-\alpha_b{(m_H^2+m_t^2)\over
2}(\Gamma^{inv}{\cal P}^{CP}+\Gamma^{CP}{\cal P}^{inv}),
\end{equation}
\begin{equation}
c_1^{inv}= 2\alpha_b{m_H^2m_t^2\over
(m_t^2-m_W^2)}\Gamma^{inv}{\cal P}^{inv},\quad
c_1^{CP}=2\alpha_b{m_H^2m_t^2\over
(m_t^2-m_W^2)}(\Gamma^{inv}{\cal P}^{CP}+\Gamma^{CP}{\cal
P}^{inv}).
\end{equation}

The asymmetry  $\Delta R^{CP}$ also measures ${\cal P}^{CP}$. We
choose $x_0={(x_{min}+x_{max})/2}$. Inserting the one-loop result
\begin{equation}
 R_{b\,\pm}= {1\over 4}\alpha_b(m_H^2-m_t^2)
\left({\cal P}^{inv}\pm {\cal P}^{CP}\right)
\label{eq:Rb}
\end{equation}
into eq.(\ref{eq:R}) gives
\begin{equation}
\label{delr}
 \Delta R^{CP}_b= {1\over 2}\alpha_b(m_H^2-m_t^2){\cal
P}^{CP}\,.
\label{eq:DeltaRbCP}
\end{equation}
%
%
\section{The \boldmath $H^{\pm}\rightarrow bb'l^{\pm} \nu$ process}  \label{sec:4}
%
%
In order to obtain the differential partial decay rate of
(\ref{pro2}), we fix the coordinate system such that the t-quark
points in the direction of the z-axis and the 3-momenta of $t$ and $l$ determine the
yz-plane:
\[ \vec{p}_{t,{\bar{t}}}=|\vec{p}_{t,{\bar{t}}}|(0,0,1), \qquad \vec{p}_{l^{\pm}}
=|\vec{p}_{l^{\pm}}|(0,\sin {\theta_{l^{\pm}}},\cos
{\theta_{l^{\pm}}}),\]
\begin{equation}
\vec{p}_{b',{\bar{b}'}}=|\vec{p}_{b',{\bar{b}'}}|(\sin
{\theta_{b',{\bar{b}'}}}\cos {\phi_{b',{\bar{b}'}}},\sin
{\theta_{b',{\bar{b}'}}}\sin {\phi_{b',{\bar{b}'}}},\cos
{\theta_{b',{\bar{b}'}}})\,.
\end{equation}
The angular distributions of $l^{\pm}$ are then given by
\begin{equation}
d\Gamma^{\pm}_l=\Gamma_{H}^{\pm}\Gamma_l^{0} \bigg [1\pm \alpha_l
m_t {(\xi _{t,\bar{t}} p_{l^{\pm}})\over (p_t p_l)}\bigg
]{E_{t,\bar{t}}\over m_t\Gamma_{t}} d\Phi_{l^{\pm}}\,,
\end{equation}
where
\begin{equation}
\Gamma_l^{0}={g^4 \pi [m_t^2-2(p_t p_l)](p_t p_l)\over 2 E_t
m_W\Gamma_W},\quad \delta (p_W^2-m_W^2)d\Phi_{l^{\pm}}={1\over
(2\pi)^4} {E_{b'}^2 E_l^2 d\Omega_{b'}d\cos \theta_l\over 4
(m_t^2-m_W^2)m_W^2}\,,
\end{equation}
\begin{equation}
(p_t p_l)=E_t E_l(1-\beta_t \cos{\theta_l})\,,
\end{equation}
\begin{equation}
E_l={m_W^2\over
2[E_t(1-\beta_t\cos{\theta_l})-E_{b'}(1-\cos{\theta_{lb'}})]}\quad
{\rm and}\quad E_{b'}={m_t^2-m_W^2\over 2E_t(1-\beta_t
\cos{\theta_{b'}})}\,.
\end{equation}
The angle $\theta_{lb'}$ is between $\vec{p}_l$ and
$\vec{p}_{b'}$:
\begin{equation}
\cos{\theta_{lb'}}=\sin {\theta_l}\sin {\theta_{b'}}\sin
{\phi_{b'}}+\cos {\theta_l}\cos {\theta_{b'}} \, .
\end{equation}
The angular distributions of $l^{\pm}$ then reads
\begin{eqnarray}
{d\Gamma^{\pm}_l\over d\cos {\theta _{l^{\pm}}}}={{\cal C}_l\over
(1-\beta_t \cos {\theta_{l^{\pm}}})^2}
 \bigg\{  \Gamma^{inv}\pm \Gamma^{CP}+ \hspace{6cm} \\ \nonumber
   \alpha_l m_t^2
 [\Gamma ^{inv}{\cal P}^{inv}\pm (\Gamma ^{inv}{\cal P}^{CP}+\Gamma ^{CP}{\cal
P}^{inv})  ] \bigg ( {E_{\bar b}(1+\cos {\theta_{l^{\pm}}} )\over
E_t (1-\beta_t\cos {\theta_{l^{\pm}})}}-{(p_tp_{\bar b})\over
m_t^2}\bigg )\bigg \}\,,
\end{eqnarray}
where
\begin{eqnarray}
{\cal C}_l=-\alpha_{\omega}^3 m_W |\vec{p}_{t}
|(m_t^2-m_W^2)^2\times \hspace{7cm} \\ \nonumber
   {[6 (1-\beta_t^2)^2 E_t^4+3 m_t^4-5 m_t^2 m_W^2+2 m_W^4-3
(1-\beta_t^2) E_t^2(3 m_t^2-2 m_W^2)]
  \over  2^8 m_H^2 m_t\Gamma_t \Gamma_W E_t^2 [m_t^2-m_W^2-(1-\beta_t^2)
  E_t^2]^3}\,.
\end{eqnarray}
$\Gamma_W$ is the total decay width of the W boson.

As there is no CP violation in $t\to bl\nu$ decay, the decay rate asymmetry
$\delta_l^{CP}$ for process
(\ref{pro2}) will measure the same combination $\Gamma^{CP}$:
\begin{equation}
  \delta_l ^{CP} =\frac{N_{l^+} - N_{l^-}}{N_{l^+} + N_{l^-}} ={\Gamma^{CP}\over
\Gamma^{inv}}\,,
\end{equation}
where $N_{l^\pm}$ are the total number of  $l^\pm$ in $H^\pm \to bb'l^\pm\nu$ decay.

For the
CP-violating FB asymmetry $\Delta A^{CP}$ of the process
(\ref{pro2}) we obtain
\begin{equation}
\label{eq:DeltaAlCP}
\Delta A^{CP}_l=2\alpha_lm_t^2m_H^2{m_H^2-m_t^2 \over(m_H^2+m_t^2)^2}{\cal P}^{CP}\,,
\end{equation}
and the FB asymmetries are at one-loop level
\begin{equation}
A^{FB}_{l\,\pm}= \beta_t+\alpha_lm_t^2m_H^2{(m_H^2-m_t^2)\over(m_H^2+m_t^2)^2}
\left({\cal P}^{inv}\pm {\cal P}^{CP}\right)\,.
\label{eq:AlFB}
\end{equation}
Notice, that the only difference between the expressions
(\ref{eq:DeltaAbCP}) and (\ref{eq:DeltaAlCP}) is the coefficient
$\alpha_b$ in (\ref{eq:DeltaAbCP}) and $\alpha_l$ in (\ref{eq:DeltaAlCP}).
These coefficients are only connected to the polarization of the t
quark. Because of the fact that $\alpha_b = 0.38$ and $\alpha_l= 1$,
one would expect a bigger effect measuring
 (\ref{eq:DeltaAlCP}).
%
%
\section{Numerical Results}  \label{sec:5}
%
Here we present a numerical analysis of the discussed asymmetries. First we analyze
the CP-violating asymmetries $\delta_{b,l}^{CP}$, $\Delta A_{b,l}^{CP}$ and $\Delta R_b^{CP}$.
Further, we study the FB asymmetries
$ A^{FB}_{b,l\,\pm}$ and $ R_{b\,\pm}$ needed for
$\Delta A_{b,l}^{CP}$ and $\Delta R_b^{CP}$.

%
\subsection{The CP-violating asymmetries}
%
The expressions ${\cal P}^{CP}$ and $\Gamma^{CP}$,
(\ref{PCP}) and (\ref{GammaCP}),
are linear combinations of the CP~violating form factors
${\rm Re}(\delta Y_t^{CP})$ and ${\rm Re}(\delta Y_b^{CP})$. Therefore,
we need to measure: 1) the
decay rate asymmetries $\delta_{b,l}^{CP}$ which are proportional to
$\Gamma^{CP}$ and 2) the angular and/or energy
asymmetries which are proportional to ${\cal P}^{CP}$.

As there is no CP violation in $t\to bW$, the decay rate asymmetries for (\ref{pro1})
and (\ref{pro2}) are equal to the decay rate
asymmetry for $H^\pm\to tb$. We denote it, following ref.\cite{we1},  by $\delta^{CP}$:
 \be
\delta^{CP}=\delta_{b}^{CP}=\delta_l^{CP}=\frac{\Gamma^{CP}}{\Gamma^{inv}}\,.
\ee

\begin{figure}[h!]
\begin{center}
\mbox{\resizebox{7.5cm}{!}{\includegraphics{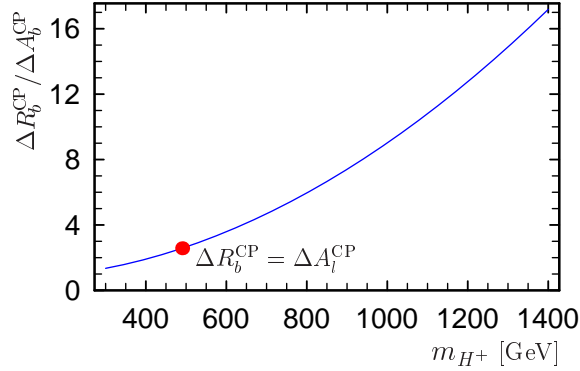}}} \vs
\end{center}
\caption[feynman]{The ratio $\Delta R_b^{\rm CP}/\Delta A_b^{\rm CP}$ as a function of $m_{H^+}$
\label{fig:1}}
\end{figure}

The angular and energy
asymmetries are not independent either. As seen from
(\ref{eq:DeltaAbCP}) and (\ref{eq:DeltaAlCP}), the angular asymmetries 
for leptons and $b$-quarks are related by:
\be
\Delta A_l^{CP} = \frac{\alpha_l}{\alpha_b}\,\, \Delta A_b^{CP} \approx 2.6\, \Delta A_b^{CP}\,.\label{r1}
\ee
Further, (\ref{eq:DeltaAbCP}) and (\ref{eq:DeltaRbCP}) lead to a simple relation between
the $b$-quark energy and angular asymmetries:
\be
\Delta R_b^{CP} = {(m_{H^+}^2 + m_t^2)^2 \over 4\, m_{H^+}^2\,
m_t^2}\, \Delta A_b^{CP}\, ,\label{r2}
\ee
which implies that for $m_{H^+} > m_t$,
$\Delta R_b^{CP}$ is bigger than $\Delta A_b^{CP}$.
Thus, in general, $\Delta R_b^{CP} $ is
the biggest asymmetry of 2) for $m_{H^+} > 490$~GeV.
Fig.~\ref{fig:1} illustrates the relative size of
the asymmetries $\Delta A_{b,l}^{CP}$ and $\Delta R_b^{CP}$ as a function of $m_{H^+}$
Note that the relations (\ref{r1}) and (\ref{r2}) are independent of $\tan\beta$.

Therefore, in the following we shall discuss only 
the decay rate asymmetry $\delta^{CP}$ and the energy asymmetry $\Delta R_b^{CP}$.
(Because of a conjugation error in our paper \cite{we1} in the formula for
the $\tilde t\tilde b\tilde g$ vertex,
see the Appendix~{\ref{sec:B}}, we have redone the numerical analysis for
$\delta^{CP}$.) The purpose of our analysis is to determine the size of the asymmetries
as functions of $m_{H^+}$ and $\tan\beta$, being the most important
parameters of the Higgs sector in MSSM,
for different values of the CP-violating phases.\\

The sources of CP violation in our processes are the one-loop
corrections to the $H^+tb$ vertex with intermediate SUSY particles, see
Fig.~1a, 1b, 1d, 1e of \cite{we1} and the self-energy graph with ${\tilde
\tau}{\tilde \nu_\tau}$. (The corrections due to Fig.~1c and
Fig.~1f of \cite{we1} are of higher order and we do not consider them here.)
In order not to deal with too many phases we assume the GUT relation
between $M_1$ and $M_2$ which fixes $\phi_{M_1}= 0$.
According to the experimental limits on the electric and
neutron EDM's, we take $\phi_\mu=0$ or $\phi_\mu=\pi /10$.
Thus, the remaining CP-violating phases in
our study are the phases of $A_t$, $A_b$ and $A_\tau$ which we shall vary.
If not specified otherwise, we fix the following values for the other MSSM parameters:
\[
    M_2=300~{\rm GeV}, \qquad M_3=745~{\rm GeV}, \qquad
     M_{\tilde U}=M_{\tilde Q}=M_{\tilde D}=M_E=M_L =350~ {\rm GeV},
\]
\begin{equation}
  \mu=-700~{\rm GeV},\qquad  |A_t| = |A_b| = |A_\tau|=700~{\rm GeV}
\label{parameters}
\end{equation}

The relevant masses of the sparticles for the
choice (\ref{parameters}) and $\tan \beta = 5$ or 30
are given in Table~\ref{table:1}.
For the case with $\phi_\mu = \pi/10$ and the other parameters
unchanged, all masses do not change by more than 1~GeV from those given in
Table~\ref{table:1}, except for $m_{\tilde t_1} = 187$~GeV and
$m_{\tilde t_2} = 515$~GeV for $\tan \beta = 5$, and $m_{\tilde t_1} = 176$~GeV for
$\tan \beta = 30$.
Note that $\phi_\mu = \pi/10$ implies  $\mu = - 700 \, e^{i \pi/10}$~GeV.
We have used running top and bottom Yukawa couplings, calculated at the
scale $Q = m_{H^+}$, see Appendix~\ref{sec:A}. We have checked that
the asymmetries have only a very weak dependence on the scale~$Q$.\\

\newcommand{\x}{@{\hspace{3mm}}}
\begin{table}
\begin{center}
\begin{tabular}{|c||c\x c\x c\x c|c\x c|c\x c|c\x c|c\x c|c|}
\hline
   $\tan\beta$
  & $m_{{\tilde \chi}^0_1}$ & $m_{{\tilde \chi}^0_2}$ & $m_{{\tilde \chi}^0_3}$ & $m_{{\tilde \chi}^0_4}$ &
   $m_{{\tilde \chi}^+_1}$ & $m_{{\tilde \chi}^+_2}$ &
   $m_{{\tilde t}_1}$ &  $m_{{\tilde t}_2}$  &  $m_{{\tilde b}_1}$  &  $m_{{\tilde b}_2}$  &
   $m_{{\tilde \tau}_1}$ & $m_{{\tilde \tau}_2}$ & $m_{{\tilde \nu}}$ \\
\hline
 5 & 142 & 300 & 706 & 706 & 300 & 709 & 166 & 522 & 327 & 377 & 344 & 362 & 344 \\
 30 & 141 & 296 & 705 & 709 & 296 & 711 & 172 & 519 & 183 & 464 & 295 & 402 & 344\\
\hline
\end{tabular}
\end{center}
\caption{Masses of the sparticles (in GeV) for the parameter set
(\ref{parameters}) together with $\phi_{A_t}=\phi_{A_b}=\pi/2$ and
$\phi_\mu$ = 0.
\label{table:1}}
\end{table}

\begin{figure}[h!]
 \begin{center}
 \mbox{\resizebox{7.5cm}{!}{\includegraphics{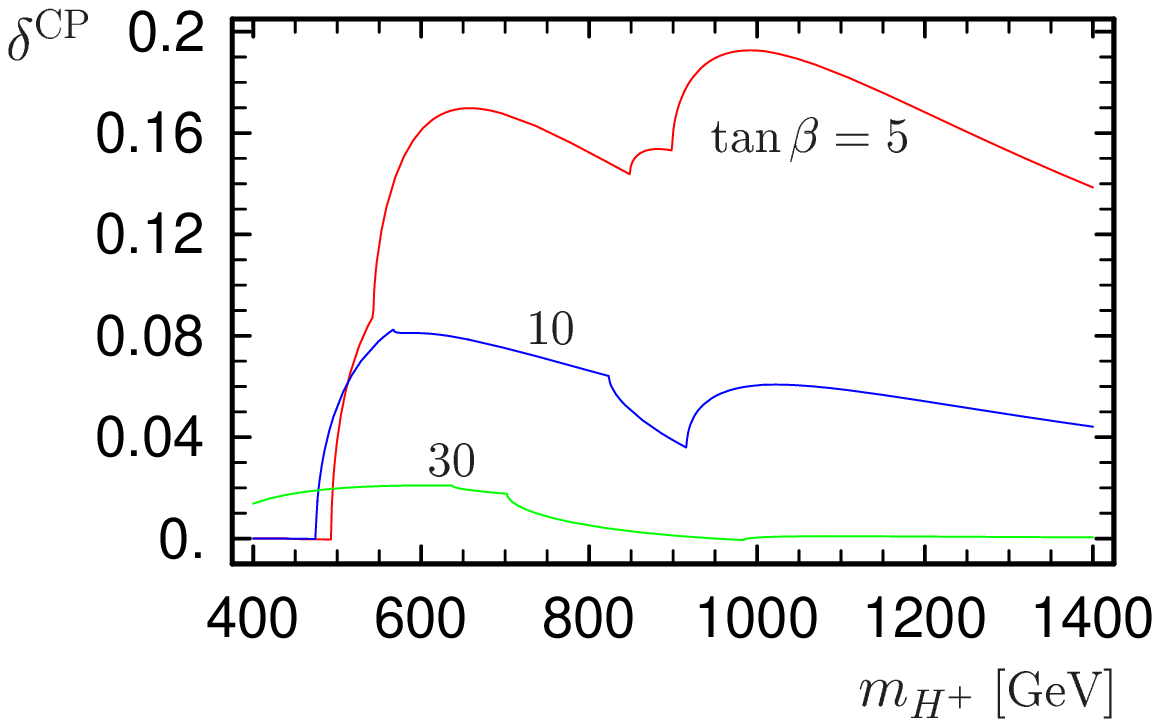}}} \hfil
 \mbox{\resizebox{7.5cm}{!}{\includegraphics{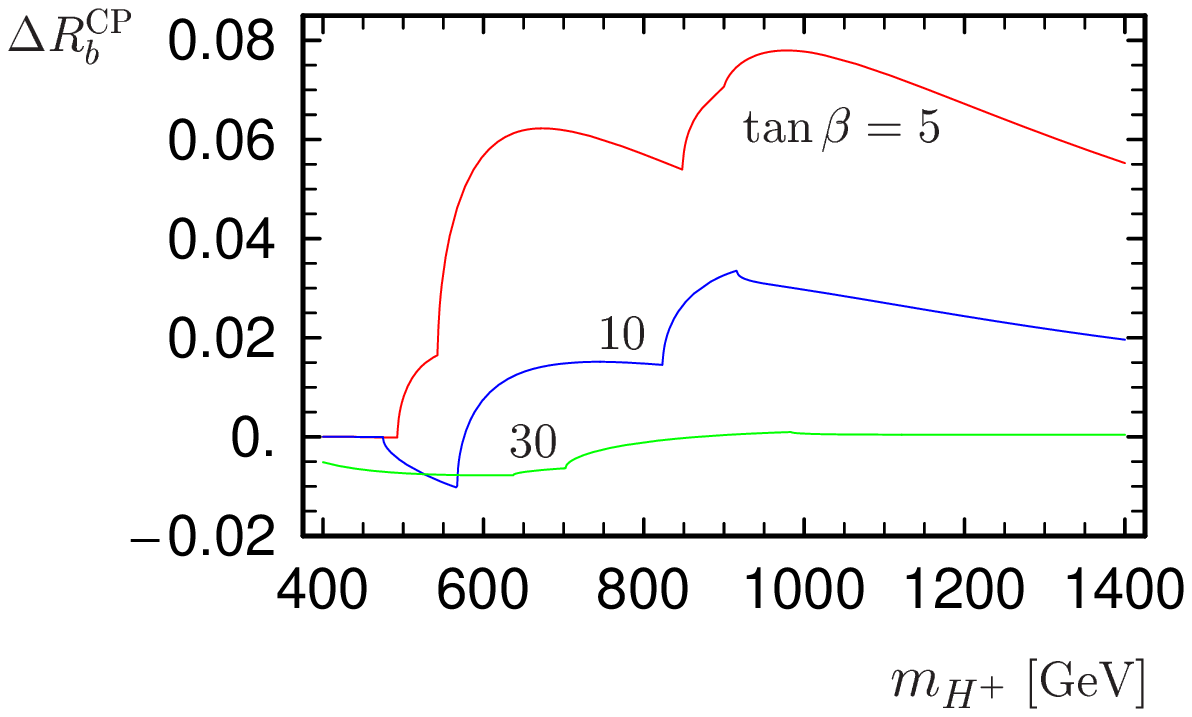}}} \vs
  \end{center}
  \caption[feynman]{The asymmetries $\delta^{CP}$ and $\Delta R_b^{CP}$ as
a function of $m_{H^+}$ for  $\phi_{A_t}=\pi/2$,\, $\phi_{A_b}=\phi_\mu =0$.
The red, blue, and green lines are for $\tan\beta=5,10$, and $30$.
\label{fig:2}}
\end{figure}

Fig.~\ref{fig:2} shows the  asymmetries $\delta^{CP}$ and $\Delta
R_b^{CP}$ as functions of $m_{H^+}$ for $\phi_{A_t}=\pi/2$, $\phi_{A_b}=0$ and
$\phi_\mu=0$. As one can see, for
$\tan\beta=5$ the decay rate asymmetry $\delta^{CP}$ goes up to $20\%$, while
$\Delta R_b^{CP}$ reaches $8\%$ for the
same values of the parameters. The asymmetries strongly depend on
$\tan\beta$ and they quickly decrease as $\tan\beta$ increases.
Our studies have shown that the most important CP-violating phase
is $\phi_{A_t}$. There is only a very weak dependence
on $\phi_{A_b}$ and $\phi_{A_\tau}$. We therefore take them zero.\\

\begin{figure}[h!]
 \begin{center}
 \mbox{\resizebox{7.5cm}{!}{\includegraphics{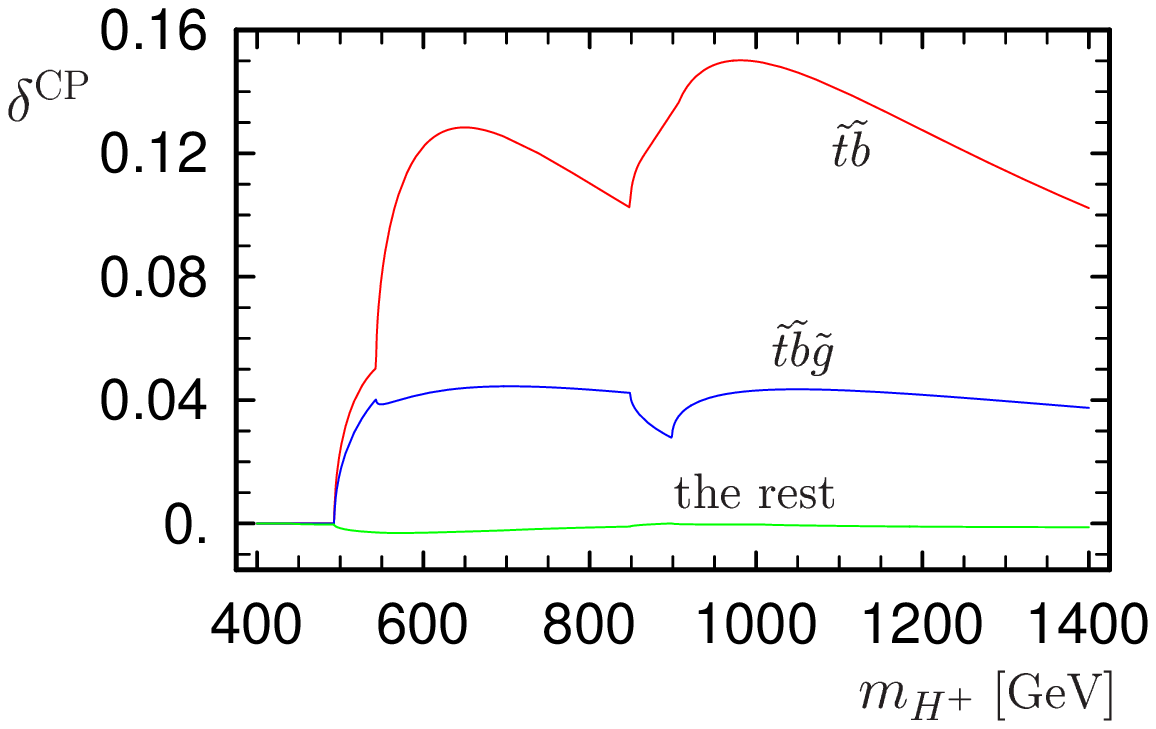}}} \hfil
 \mbox{\resizebox{7.5cm}{!}{\includegraphics{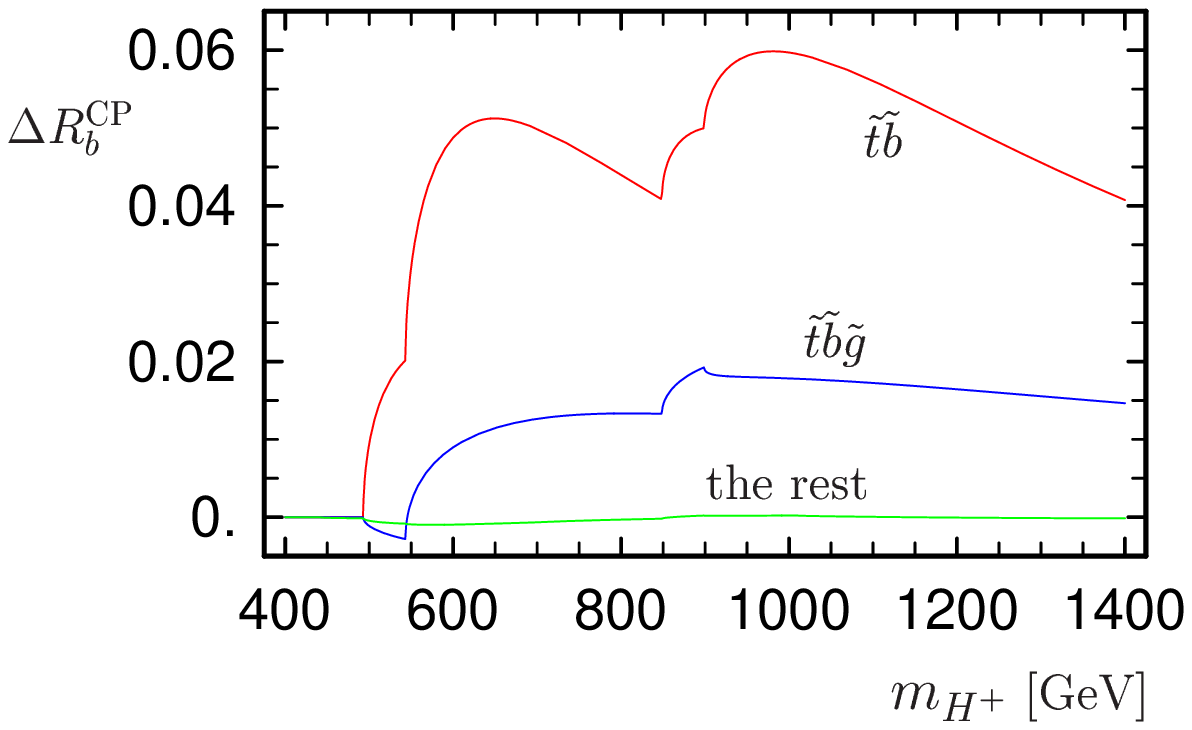}}} \vs
  \end{center}
  \caption[feynman]{The contribution of the $\tilde t\tilde b$ self-energy
 (red line), $\tilde t\tilde b\tilde g$ vertex contribution (blue line) and
the sum of the other (green line) diagrams to
$\delta^{CP}$ and $\Delta R_b^{CP}$ as a function of $m_{H^+}$
for $\tan\beta =5$ and $\phi_{A_t}=\pi/2$, $\phi_{A_b}=\phi_\mu =0$.
\label{fig:3}}
\end{figure}

The main contributions to both $\delta^{CP}$ and $\Delta R_b^{CP}$
come from the self-energy graph with stop-sbottom. The vertex graph with
stop-sbottom-gluino also gives a non-zero contribution. Their
contributions are shown in Fig.~\ref{fig:3}. The contribution of
the rest of the graphs is negligible. This justifies the use of the GUT relation which
fixes $\phi_{M_1}=0$, and it explains the weak dependence on $\phi_{A_\tau}$.\\

\begin{figure}[h!]
 \begin{center}
 \mbox{\resizebox{7.5cm}{!}{\includegraphics{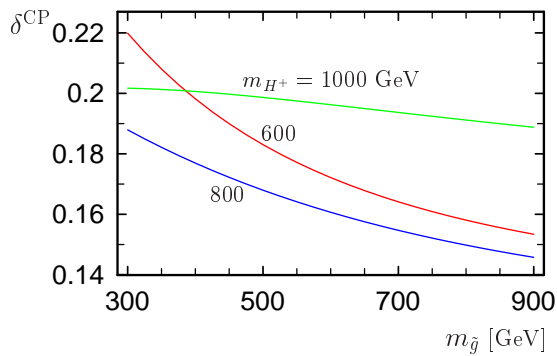}}} \vs
  \end{center}
  \caption[feynman]{$\delta^{CP}$ as a function of the gluino mass
  for $\tan\beta=5$. The red line is for $m_{H^+}=600 $ GeV, the blue line is for
  $m_{H^+}=800$ GeV and the green line is for $m_{H^+}=1000 $ GeV.
\label{fig:4}}
\end{figure}

Up to now, all the analyses are done for
$M_3=m_{\tilde{g}}=744~{\rm GeV}$. In Fig.~\ref{fig:4}, we show the dependence
of $\delta^{CP}$ on the gluino mass. In general, $\delta^{CP}$ gets small
with increasing gluino mass.

\begin{figure}[h!]
 \begin{center}
 \mbox{\resizebox{7.5cm}{!}{\includegraphics{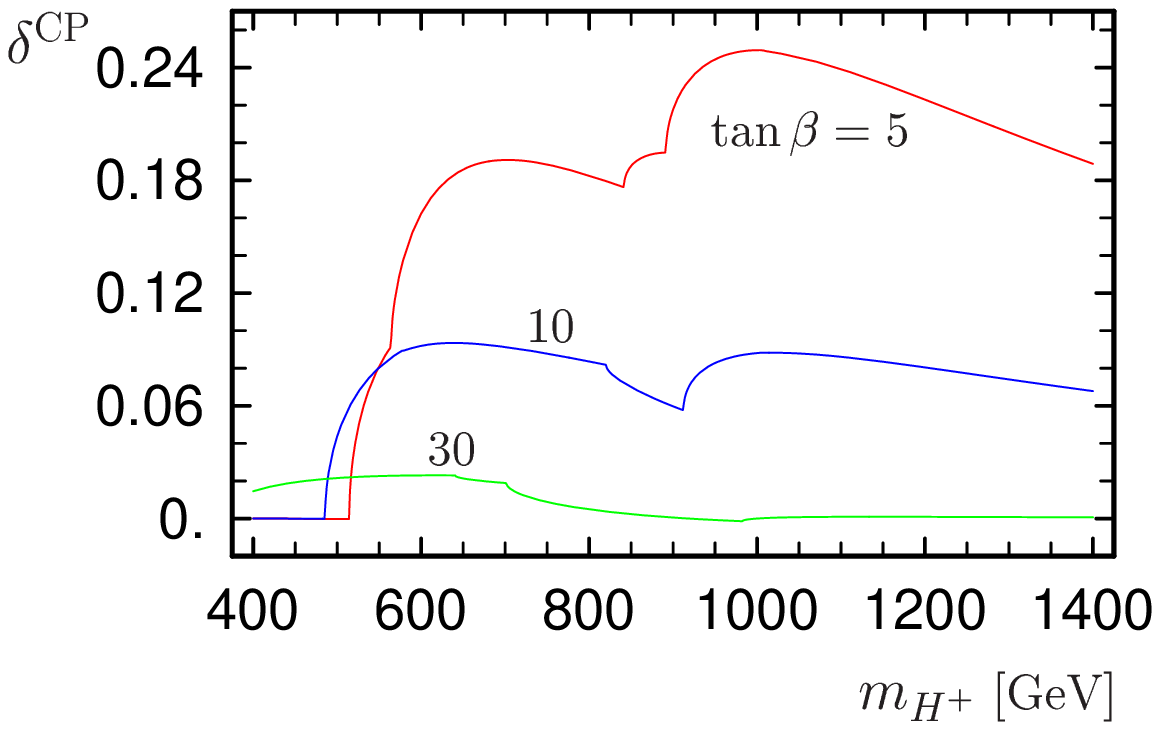}}} \hfill
 \mbox{\resizebox{7.5cm}{!}{\includegraphics{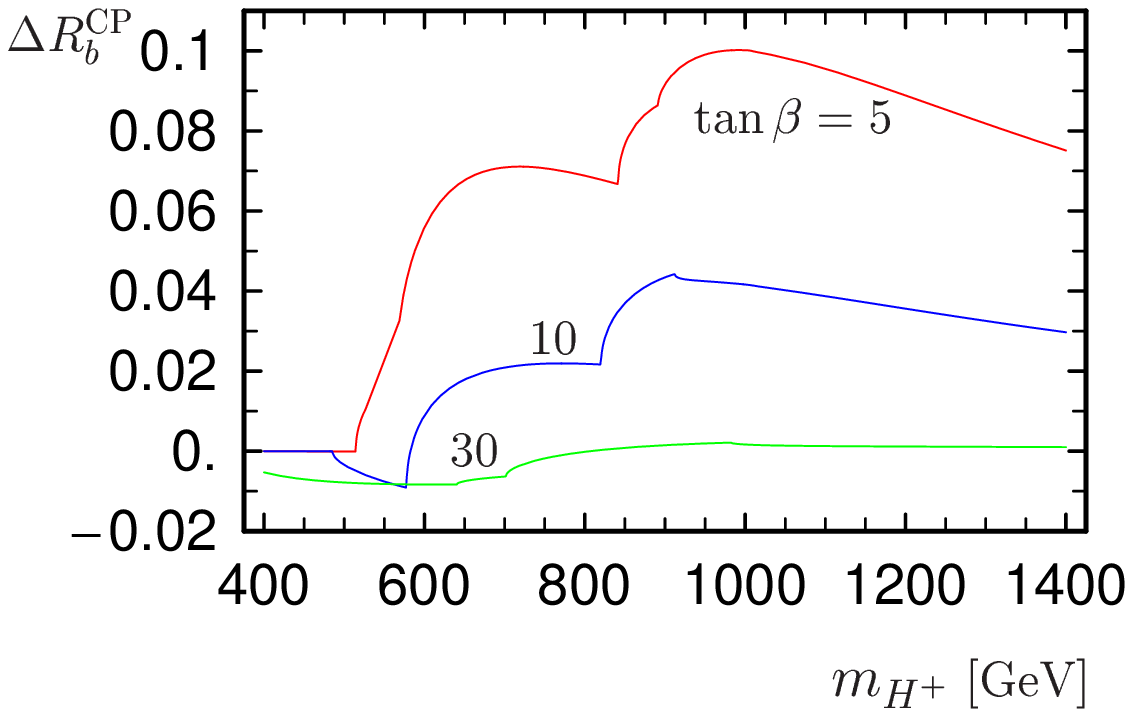}}} \vs
  \end{center}
  \caption[feynman]{The asymmetries $\delta^{CP}$ and $\Delta R_b^{CP}$ as
a function of $m_{H^+}$ for  $\phi_{A_t}=\pi/2$, $\phi_{A_b}=0$ and a non
zero phase of $\mu$, $\phi_\mu=\pi/10$. The red, blue and green lines are for
$\tan\beta=5,10$, and $30$.
\label{fig:5}}
\end{figure}

Let us now allow a non-zero phase of $\mu$. We take a very small phase,
$\phi_\mu = \pi/10$, in order
not to be in contradiction with the experimental data.
As can be seen in Fig.~\ref{fig:5}, the asymmetries can increase up to $25\%$
for $\delta^{\rm CP}$ and $10\%$ for $\Delta R_b^{\rm CP}$.
The discussed asymmetries $\delta^{CP}$ and $\Delta R_b^{CP }$ show a
very strong dependence on
the sign of $\mu$. As noted above, our analysis is done for $\mu = - 700$ (see (\ref{parameters})),
however if $\mu$ changes sign, $\mu =700$, all asymmetries become extremely small.

%
%
\subsection{The P-violating asymmetries}
%
When discussing the possibilities to measure $\Delta A_{b,l}^{CP}$ and $\Delta R_b^{CP}$,
it is also important to know the size of the FB asymmetries $A_{b,l\,\pm}^{FB}$,
(\ref{eq:AbFB}) and (\ref{eq:AlFB}), and of
the energy asymmetry $R_{b\,\pm}$, (\ref{eq:Rb}),
that enter the corresponding CP-violating asymmetries.\\

$A_{b,l\,\pm}^{FB}$ and $R_{b\, \pm}$ are determined by the
polarization ${\cal P}^\pm$ of the $t$-quark in $H^\pm \to t b$ decays.
As the Lagrangian violates parity, these asymmetries appear already
at tree level and thus should be rather large.\\

Neglecting the loop induced CP-violating part ${\cal P}^{CP}$
in (\ref{eq:AbFB}), (\ref{eq:AlFB}), and (\ref{eq:Rb}), we get
\be
A_{b,l}^{inv} = \frac12 \left(A_{b,l\,+}^{FB} + A_{b,l\,-}^{FB} \right)
&=&\beta_t + \alpha_{b,l}\,m_t^2 m_{H^+}^2 {(m_{H^+}^2-m_t^2)\over(m_{H^+}^2+m_t^2)^2}\,{\cal
P}^{inv} \, ,\\
R_{b}^{inv} = \frac12 \left(R_{b\,+} + R_{b\,-}\right) &=& {1\over 4}\alpha_b(m_{H^+}^2-m_t^2)\,{\cal P}^{inv} \simeq
{\alpha_b \over 2} {y_t^2 - y_b^2 \over y_t^2 + y_b^2}  \, .
\ee
Thus $\Delta A_{b,l}^{CP}$, (\ref{eq:DeltaAbCP}, \ref{eq:DeltaAlCP}),
and $\Delta R_b^{CP}$, (\ref{eq:DeltaAlCP}), are determined by ${\cal P}^{CP}$, while
$A_{b,l}^{FB}$ and $R_b$ are determined by ${\cal P}^{inv}$, and there is no a
priori reason to expect that their
$\tan\beta$ and $m_{H^+}$ dependences will be the same.
$A_{b,l}^{inv}$ and $R_{b}^{inv}$ are tree-level quantities. Including the one-loop corrections
to these would require the full renormalization of the process which is beyond
the scope of this paper.\\

\begin{figure}[h!]
 \begin{center}
 \mbox{\resizebox{7.5cm}{!}{\includegraphics{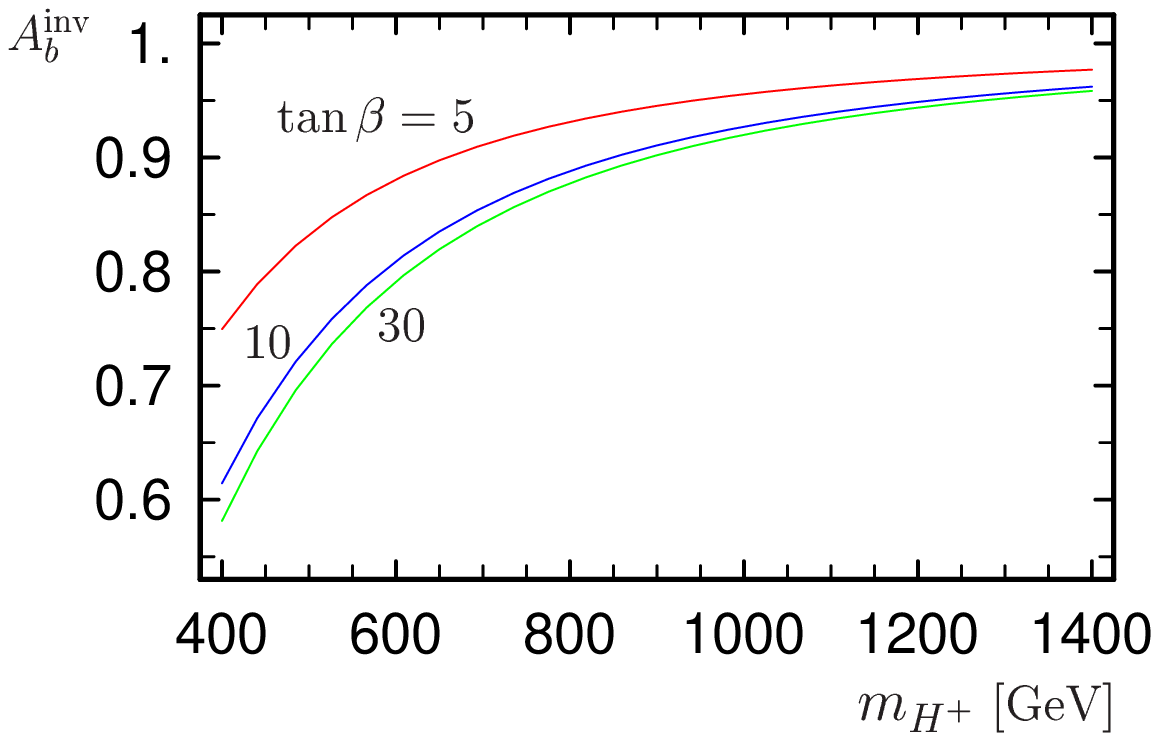}}} \hfill
 \mbox{\resizebox{7.5cm}{!}{\includegraphics{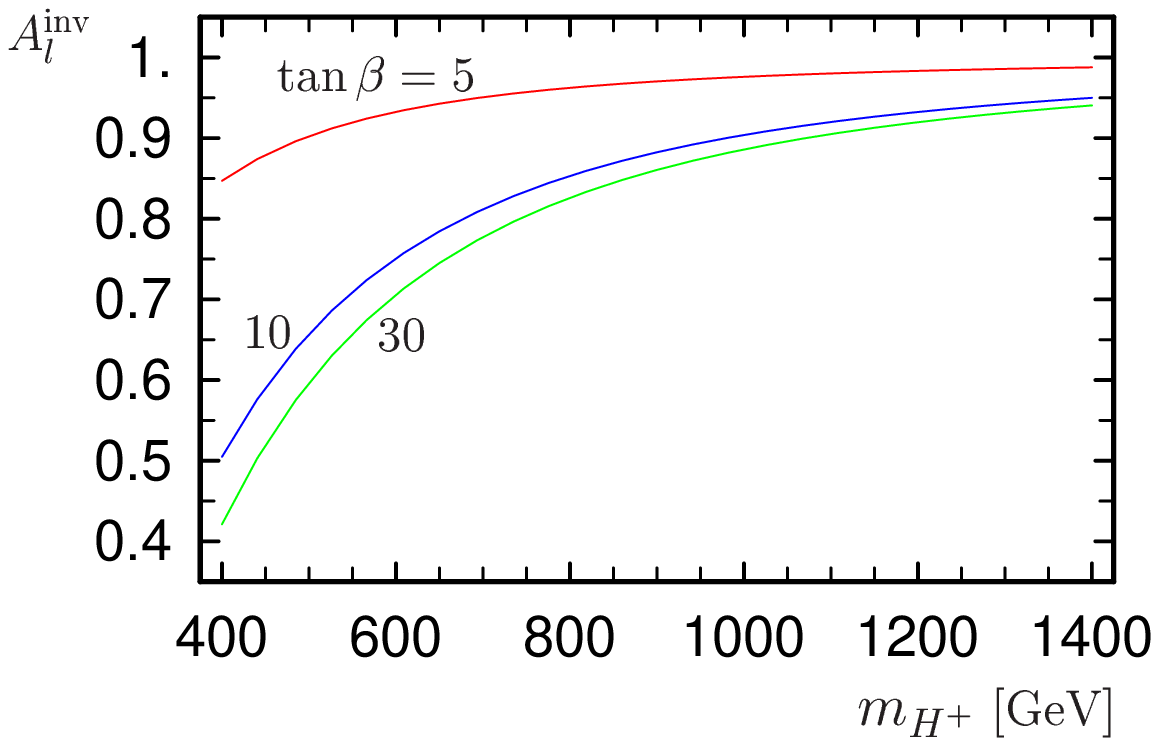}}} \vs
  \end{center}
  \caption[feynman]{The forward-backward asymmetries $A_b^{inv}$ and $A_l^{inv}$
as a function of $m_{H^+}$ for $\tan\beta = 5$~(red), 10~(blue), and 30~(green).
\label{fig:7}}
\end{figure}

\begin{figure}[h!]
 \begin{center}
 \mbox{\resizebox{7.5cm}{!}{\includegraphics{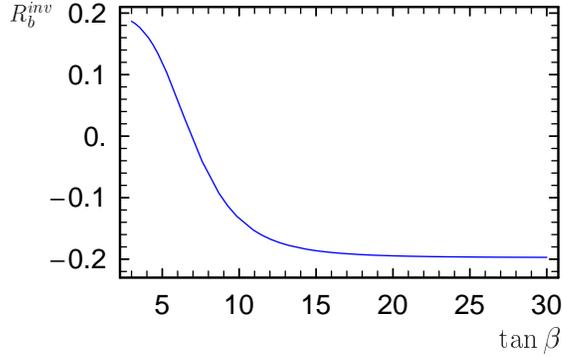}}} \vs
  \end{center}
  \caption[feynman]{The energy asymmetry $R_b^{inv}$ as a function of $\tan\beta$
\label{fig:8}}
\end{figure}

In Fig.~\ref{fig:7} we present
$A_b^{inv}$ and $A_l^{inv}$ as a function of $m_{H^+}$ for $\tan\beta$ = 5,10, and 30,
and in Fig.~\ref{fig:8} we show $ R_b^{inv}$ as a function of $\tan\beta$.
Note that at tree-level $A_{b,l}^{FB}$ depends only on $\tan\beta$ and $m_{H^+}$,
and $R_b$ only on $\tan\beta$.

%
\section{Conclusions} \label{sec:6}
%
We have calculated the CP-violating decay rate, forward-backward and
energy asymmetries between $H^{+}\rightarrow \bar{b}\,t\rightarrow
\bar{b}\,b'\,W^{+} ( \rightarrow \bar{b}\,b'\,l^{+}\,\nu_l)$ and
$H^{-}\rightarrow b\,\bar{t}\rightarrow b\,\bar{b}'\,W^{-}
(\rightarrow b\,\bar{b}'\, l^{-}\,\nu_l)$. They are induced by loop
corrections in the $H^{\pm} t b$- vertex. The CP~violating 
forward-backward and energy asymmetries are determined
by the polarization of the top quark and are therefore related.
We have shown that it is necessary to measure {\it both} the
decay rate asymmetry $\delta^{CP}$ {\it and} the forward-backward or
the energy asymmetry to get the maximal information on the
CP-violating parts of the decay amplitude.
We have performed a detailed numerical analysis of these quantities.
An important improvement with running Yukawa couplings at the $m_{H^{+}}$ scale
has been made. The asymmetries are most sensitive to the phase 
$\phi_{A_t}$. The asymmetries reach their maximum for $\tan\beta=5$ and $\mu=-700$~GeV.
The decay rate asymmetry can go up to $25\%$, the others up to $10\%$.
The main contribution comes from the self-energy
diagram with stop and sbottom exchange. 
We have also calculated the P-violating asymmetries at tree level.\\

We want to add a few remarks on the measurability of these asymmetries.
In principle, the production rate for $H^{\pm}$ at LHC is not so
small being 0.2 pb for $m_{H^+} =500 $ GeV and $\tan \beta =30$
\cite{Belyaev, barger}. The main production process is due to $g\
{\bar b}\ \rightarrow\  H^+\ {\bar t}$. Because of the large
background, the actual signal production rate is strongly reduced.
According to \cite{Belyaev}, one can expect $N=733$ signals with
$N/\sqrt{B}=12.6$ for $m_{H^+} =500 $ GeV, $\tan \beta =50$ for a
luminosity ${\cal L}=100\ {\rm fb}^{-1}$. The statistical
significance $\sqrt{N} $ A to measure an asymmetry A of several
percent might be too low for a clear observation of CP violation
in  $H^{+}$ decays at LHC in the first stage. However, at SLHC for
which a luminosity of 1000-3000 ${\rm fb}^{-1}$ is designed, such a
measurement would be worth of being performed. Here we have only
considered CP violation in the $H^{+}$ decays. However, similar
graphs are also present in the production process $g\ {\bar b}\
\rightarrow\  H^+\ {\bar t}$~\cite{jennifer}. One would expect a CP-violating
asymmetry of similar size. The total asymmetry in production and
decay would be approximately additive,
$A_{\rm tot.}= A_{\rm prod.}+ A_{\rm decay}$.

\section*{Acknowledgements}
%
We thank Jennifer Williams for finding the conjugation error in \cite{we1}.
The authors acknowledge support from EU under the MRTN-CT-2006-035505 network
programme. This work is supported by the "Fonds zur F\"orderung der
wissenschaftlichen Forschung" of Austria, project No. P18959-N16.
%
%
%
\clearpage
\begin{appendix}
\section{Running Yukawa couplings} \label{sec:A}
For clarity, we present all formulas used for programming the
running top and bottom Yukawa couplings, $h_b$ and $h_t$, respectively.
The Lagrangian for the $H^\pm t b$ interactions reads
\begin{equation}
{\cal L}_{Hqq} =
  H^+\,\bar{t}\,(y_b^*\PR + y_t\PL)\,b + H^-\,\bar{b}\,(y_t^*\PR + y_b\PL)\,t
  + \ldots \,,
\label{eq:A10}
\end{equation}
with the $\drbar$ running top and bottom Yukawa couplings in the MSSM,
\begin{equation}
  y_b = h_b\sin\beta \,,\qquad
  y_t = h_t\cos\beta\,,
\end{equation}
given at the scale $Q = m_{H^+}$ in our studied case, and
\begin{equation}
  h_b  =   {g m^{\drbar}_{b}(Q) \over \sqrt{2} m_W \cos\beta}\, , \qquad
  h_t  =   {g m^{\drbar}_{t}(Q) \over \sqrt{2} m_W \sin\beta}\, .
\end{equation}
In \cite{carenaCP} it is shown that within an effective theory approach
large finite scale independent parts can be resummed, which in case
of complex MSSM input parameters leads  to complex $h_b$ and $h_t$. Effective means
that the masses of the particles in the loops are much bigger than those
of the in- and outgoing
particles so that these states can be integrated out in the Lagrangian.
In our case, we are interested in additional open channels, e.~g.
$H^+ \to \tilde t \bar{\tilde b}$. This implies that the
resummation is not applicable here.
But we can improve our calculation by using full one-loop running quark masses
with some higher order improvements of the gluonic part.
Note, that $m^{\drbar}_{q}(Q)$ can always be made real
by field redefinition \cite{carenaCP} and therefore also $h_q$ is real
in our case.

We take as input set the bottom mass
$m^{\msbar}_b(m_b) = $~4.2~GeV, the mass for the top quark
is the pole mass, $m^{\rm pole}_t =$~171.4~GeV, the strong coupling is
$\alpha^{\msbar}_s(m_Z) =$~0.1176, $m_Z =$~91.1876~GeV, and $m_W =$~80.406~GeV

First we want to have the $\drbar$ bottom mass of the Standard Model at the scale
$Q$ (see for comparison eq.~(26) in \cite{SPApaper}),
 \begin{eqnarray}
  m^{\drbar}_{b,\,\rm SM}(Q) &=& m_b^{\msbar}(Q)
  \left[ 1 - \frac{\alpha^\drbar_s}{3 \pi}
    - \frac{23 \alpha^{2,\drbar}_s}{72 \pi} \right]
\end{eqnarray}
with $m_b^{\msbar}(Q) \equiv m_b(Q)_{\rm SM}$ given in \cite{improved,improved-ref14}.
Adding the loop contributions due to supersymmetric and
heavy SM particles, denoted by $\Delta m_b^{\rm extra}$
(calculated in the $\drbar$ renormalization scheme),
we get the full one-loop $\drbar$ running bottom mass
(with some higher-order improvements) within the MSSM,
\begin{equation}
  m^{\drbar}_{b}(Q) = m^{\drbar}_{b,\,\rm SM}(Q) + \Delta m_b^{\rm extra}(Q) \, .
\label{eq:mbrunDRbar}
\end{equation}

The $\drbar$ running top mass we get from
\begin{eqnarray}
  m_t^{\drbar}(Q) &=& m^{\rm pole}_t
  \Bigg[ 1 + {\Delta m_t^{(1)} \over m_t} + {\Delta m_t^{(2, g)} \over m_t} \Bigg]
\end{eqnarray}
where $\Delta m_t^{(1)}$ is the full one-loop contribution
to $m_t$ (calculated in the $\drbar$ renormalization scheme)
and $\Delta m_t^{(2, g)}$ is the gluonic two-loop contribution,
\begin{equation}
  {\Delta m_t^{(2, g)} \over m_t} = - \left({\alpha_s(Q) \over 4 \pi}\right)^2
\left(\frac{8\,{\pi }^2}{9} +  \frac{2011}{18}+ \frac{16\,{\pi }^2\,\log (2)}{9}  -
  \frac{8\,{\xi}(3)}{3} + 82\,L + 22\,L^2    \right)\, ,
\label{eq:delmt2loopg}
\end{equation}
with $L = \log(Q^2/m^2_t)$, see \cite{mt2loopg}.

%
\section{Squark--quark--gluino contribution} \label{sec:B}
In Appendix~B of \cite{we1}, eq.~(62) is incorrect and therefore
also eqs.~(14, 15). For clarification, the definition of the squark
rotation matrix $R^{\,\sq}$ is essential.
In this work and in \cite{we1}, one has
\begin{equation}
\tilde{q}_{\alpha}=R^*_{i\alpha}\tilde{q}_i\quad {\rm with}\quad
\alpha=L,R \quad {\rm and}\quad i=1,2\,.
\end{equation}

Hence, the
squark--quark--gluino interaction ( eq.~(62) of \cite{we1} ) is
given by
\begin{eqnarray}
  {\cal L}_{q\sq\sg} &=&
  -\rzw\,g_s\,T_{st}^a\left[\,
      \bar{\sg}^a (R_{1i}^{\sq*}\,e^{-\frac{i}{2}\phi_3}\PL -
                   R_{2i}^{\sq*}\,e^{ \frac{i}{2}\phi_3}\PR)\,
      q_s^{}\,\sq_{i,t}^* \right.\nn\\
  & & \hspace{3cm}\left.
      +\,\bar q_s^{} (R_{1i}^{\sq}\,e^{ \frac{i}{2}\phi_3}\PR -
                   R_{2i}^{\sq}\,e^{-\frac{i}{2}\phi_3}\PL)\,\sg^a\,\sq_{i,t}
      \,\right]\,,
\end{eqnarray}

The contribution from the diagram with a stop, a sbottom, and a
gluino (in \cite{we1} eqs.~(14,15)) is
\begin{eqnarray}
  {\rm Re
}\,\delta Y^{CP}_b(\st_i\,\sb_j\sg) & = &
-\frac{4}{3}\frac{\alpha_s}{\pi}\,
    \big[ \,
      m_\sg\,{\rm Im}(G_{4ij}\Rst_{1i}\Rsbs_{2j}\,e^{i\phi_3})\,{\rm Im}(C_0) \nn\\
    && \hspace{-5mm}
      + m_t\,{\rm Im}(G_{4ij}\Rst_{2i}\Rsbs_{2j})\,{\rm Im}(C_1)
      + m_b\,{\rm Im}(G_{4ij}\Rst_{1i}\Rsbs_{1j})\,{\rm Im}(C_2) \,\big]\,,\\
 {\rm Re}\,\delta Y^{CP}_t(\st_i\,\sb_j\sg) & = & -\frac{4}{3}\frac{\alpha_s}{\pi}\,
    \big[ \,
      m_\sg\,{\rm Im}(G_{4ij}\Rst_{2i}\Rsbs_{1j}\,e^{-i\phi_3})\,{\rm Im}(C_0) \nn\\
    && \hspace{-5mm}
      + m_t\,{\rm Im}(G_{4ij}\Rst_{1i}\Rsbs_{1j})\,{\rm Im}(C_1)
      + m_b\,{\rm Im}(G_{4ij}\Rst_{2i}\Rsbs_{2j})\,{\rm Im}(C_2) \,\big]\,,
      \end{eqnarray}
with $C_X=C_X(m_t^2,m_{H^+}^2,m_b^2,m_\sg^2,m_{\st_i}^2,m_{\sb_j}^2)$.
\end{appendix}


\end{document}